\magnification 1200

%
%
\newdimen\FigSize       \FigSize=.9\hsize 
%
\newskip\abovefigskip   \newskip\belowfigskip
\gdef\epsfig#1;#2;{\par\vskip\abovefigskip\penalty -500
   {\everypar={}\epsfxsize=#1\nd
    \centerline{\epsfbox{#2}}}%
    \vskip\belowfigskip}%
%
\newskip\figtitleskip
\gdef\tepsfig#1;#2;#3{\par\vskip\abovefigskip\penalty -500
   {\everypar={}\epsfxsize=#1\nd
    \vbox
      {\centerline{\epsfbox{#2}}\vskip\figtitleskip
       \centerline{\figtitlefont#3}}}%
    \vskip\belowfigskip}%
%
\newcount\FigNr \global\FigNr=0
\gdef\nepsfig#1;#2;#3{\global\advance\FigNr by 1
   \tepsfig#1;#2;{Figure\space\the\FigNr.\space#3}}%
%
%
%
\gdef\ipsfig#1;#2;{
   \midinsert{\everypar={}\epsfxsize=#1\nd
              \centerline{\epsfbox{#2}}}%
   \endinsert}%
%
\gdef\tipsfig#1;#2;#3{\midinsert
   {\everypar={}\epsfxsize=#1\nd
    \vbox{\centerline{\epsfbox{#2}}%
          \vskip\figtitleskip
          \centerline{\figtitlefont#3}}}\endinsert}%
%
\gdef\nipsfig#1;#2;#3{\global\advance\FigNr by1%
  \tipsfig#1;#2;{Figure\space\the\FigNr.\space#3}}%
\newread\epsffilein    
\newif\ifepsffileok    
\newif\ifepsfbbfound   
\newif\ifepsfverbose   
\newdimen\epsfxsize    
\newdimen\epsfysize    
\newdimen\epsftsize    
\newdimen\epsfrsize    
\newdimen\epsftmp      
\newdimen\pspoints     
\pspoints=1bp          
\epsfxsize=0pt         
\epsfysize=0pt         
\def\epsfbox#1{\global\def\epsfllx{72}\global\def\epsflly{72}%
   \global\def\epsfurx{540}\global\def\epsfury{720}%
   \def\lbracket{[}\def\testit{#1}\ifx\testit\lbracket
   \let\next=\epsfgetlitbb\else\let\next=\epsfnormal\fi\next{#1}}%
\def\epsfgetlitbb#1#2 #3 #4 #5]#6{\epsfgrab #2 #3 #4 #5 .\\%
   \epsfsetgraph{#6}}%
\def\epsfnormal#1{\epsfgetbb{#1}\epsfsetgraph{#1}}%
\def\epsfgetbb#1{%
%
%
\openin\epsffilein=#1
\ifeof\epsffilein\errmessage{I couldn't open #1, will ignore it}\else
%
%
   {\epsffileoktrue \chardef\other=12
    \def\do##1{\catcode`##1=\other}\dospecials \catcode`\ =10
    \loop
       \read\epsffilein to \epsffileline
       \ifeof\epsffilein\epsffileokfalse\else
%
%
          \expandafter\epsfaux\epsffileline:. \\%
       \fi
   \ifepsffileok\repeat
   \ifepsfbbfound\else
    \ifepsfverbose\message{No bounding box comment in #1; using
defaults}\fi\fi
   }\closein\epsffilein\fi}%
%
%
\def\epsfsetgraph#1{%
   \epsfrsize=\epsfury\pspoints
   \advance\epsfrsize by-\epsflly\pspoints
   \epsftsize=\epsfurx\pspoints
   \advance\epsftsize by-\epsfllx\pspoints
%
%
   \epsfxsize\epsfsize\epsftsize\epsfrsize
   \ifnum\epsfxsize=0 \ifnum\epsfysize=0
      \epsfxsize=\epsftsize \epsfysize=\epsfrsize
%
arithmetic!
%
     \else\epsftmp=\epsftsize \divide\epsftmp\epsfrsize
       \epsfxsize=\epsfysize \multiply\epsfxsize\epsftmp
       \multiply\epsftmp\epsfrsize \advance\epsftsize-\epsftmp
       \epsftmp=\epsfysize
       \loop \advance\epsftsize\epsftsize \divide\epsftmp 2
       \ifnum\epsftmp>0
          \ifnum\epsftsize<\epsfrsize\else
             \advance\epsftsize-\epsfrsize \advance\epsfxsize\epsftmp
\fi
       \repeat
     \fi
   \else\epsftmp=\epsfrsize \divide\epsftmp\epsftsize
     \epsfysize=\epsfxsize \multiply\epsfysize\epsftmp
     \multiply\epsftmp\epsftsize \advance\epsfrsize-\epsftmp
     \epsftmp=\epsfxsize
     \loop \advance\epsfrsize\epsfrsize \divide\epsftmp 2
     \ifnum\epsftmp>0
        \ifnum\epsfrsize<\epsftsize\else
           \advance\epsfrsize-\epsftsize \advance\epsfysize\epsftmp \fi
     \repeat
   \fi
%
%
   \ifepsfverbose\message{#1: width=\the\epsfxsize,
height=\the\epsfysize}\fi
   \epsftmp=10\epsfxsize \divide\epsftmp\pspoints
   \vbox to\epsfysize{\vfil\hbox to\epsfxsize{%
      \includegraphics{#1}%
      \hfil}}%
\epsfxsize=0pt\epsfysize=0pt}%
%
%
{\catcode`\%=12
\global\let\epsfpercent=
%
%
\long\def\epsfaux#1#2:#3\\{\ifx#1\epsfpercent
   \def\testit{#2}\ifx\testit\epsfbblit
      \epsfgrab #3 . . . \\%
      \epsffileokfalse
      \global\epsfbbfoundtrue
   \fi\else\ifx#1\par\else\epsffileokfalse\fi\fi}%
%
%
\def\epsfgrab #1 #2 #3 #4 #5\\{%
   \global\def\epsfllx{#1}\ifx\epsfllx\empty
      \epsfgrab #2 #3 #4 #5 .\\\else
   \global\def\epsflly{#2}%
   \global\def\epsfurx{#3}\global\def\epsfury{#4}\fi}%
%
%
\def\epsfsize#1#2{\epsfxsize}%
%
%

\epsfverbosetrue                        
\abovefigskip=\baselineskip             
\belowfigskip=\baselineskip             
\global\let\figtitlefont\bf             
\global\figtitleskip=.5\baselineskip    

\font\tenmsb=msbm10   
\font\sevenmsb=msbm7
\font\fivemsb=msbm5
\newfam\msbfam
\textfont\msbfam=\tenmsb
\scriptfont\msbfam=\sevenmsb
\scriptscriptfont\msbfam=\fivemsb
\def\Bbb#1{\fam\msbfam\relax#1}
\let\nd\noindent 

\def\natural{{\rm I\kern-.18em N}}
\newskip\ttglue


\def\eightpoint{\def\rm{\fam0\eightrm}  
  \textfont0=\eightrm \scriptfont0=\sixrm \scriptscriptfont0=\fiverm
  \textfont1=\eighti  \scriptfont1=\sixi  \scriptscriptfont1=\fivei
  \textfont2=\eightsy  \scriptfont2=\sixsy  \scriptscriptfont2=\fivesy
  \textfont3=\tenex  \scriptfont3=\tenex  \scriptscriptfont3=\tenex
  \textfont\itfam=\eightit  \def\it{\fam\itfam\eightit}
  \textfont\slfam=\eightsl  \def\sl{\fam\slfam\eightsl}
  \textfont\ttfam=\eighttt  \def\tt{\fam\ttfam\eighttt}
  \textfont\bffam=\eightbf  \scriptfont\bffam=\sixbf
    \scriptscriptfont\bffam=\fivebf  \def\bf{\fam\bffam\eightbf}
  \tt  \ttglue=.5em plus.25em minus.15em
  \normalbaselineskip=9pt
  \setbox\strutbox=\hbox{\vrule height7pt depth2pt width0pt}
  \let\sc=\sixrm  \let\big=\eightbig \normalbaselines\rm}

\font\eightrm=cmr8 \font\sixrm=cmr6 \font\fiverm=cmr5
\font\eighti=cmmi8  \font\sixi=cmmi6   \font\fivei=cmmi5
\font\eightsy=cmsy8  \font\sixsy=cmsy6 \font\fivesy=cmsy5
\font\eightit=cmti8  \font\eightsl=cmsl8  \font\eighttt=cmtt8
\font\eightbf=cmbx8  \font\sixbf=cmbx6 \font\fivebf=cmbx5

\def\eightbig#1{{\hbox{$\textfont0=\ninerm\textfont2=\ninesy
        \left#1\vbox to6.5pt{}\right.\enspace$}}}

\def\ninepoint{\def\rm{\fam0\ninerm}  
  \textfont0=\ninerm \scriptfont0=\sixrm \scriptscriptfont0=\fiverm
  \textfont1=\ninei  \scriptfont1=\sixi  \scriptscriptfont1=\fivei
  \textfont2=\ninesy  \scriptfont2=\sixsy  \scriptscriptfont2=\fivesy
  \textfont3=\tenex  \scriptfont3=\tenex  \scriptscriptfont3=\tenex
  \textfont\itfam=\nineit  \def\it{\fam\itfam\nineit}
  \textfont\slfam=\ninesl  \def\sl{\fam\slfam\ninesl}
  \textfont\ttfam=\ninett  \def\tt{\fam\ttfam\ninett}
  \textfont\bffam=\ninebf  \scriptfont\bffam=\sixbf
    \scriptscriptfont\bffam=\fivebf  \def\bf{\fam\bffam\ninebf}
  \tt  \ttglue=.5em plus.25em minus.15em
  \normalbaselineskip=11pt
  \setbox\strutbox=\hbox{\vrule height8pt depth3pt width0pt}
  \let\sc=\sevenrm  \let\big=\ninebig \normalbaselines\rm}

\font\ninerm=cmr9 \font\sixrm=cmr6 \font\fiverm=cmr5
\font\ninei=cmmi9  \font\sixi=cmmi6   \font\fivei=cmmi5
\font\ninesy=cmsy9  \font\sixsy=cmsy6 \font\fivesy=cmsy5
\font\nineit=cmti9  \font\ninesl=cmsl9  \font\ninett=cmtt9
\font\ninebf=cmbx9  \font\sixbf=cmbx6 \font\fivebf=cmbx5
\def\ninebig#1{{\hbox{$\textfont0=\tenrm\textfont2=\tensy
        \left#1\vbox to7.25pt{}\right.$}}}

\font\tenmsb=msbm10   
\font\sevenmsb=msbm7
\font\fivemsb=msbm5
\newfam\msbfam
\textfont\msbfam=\tenmsb
\scriptfont\msbfam=\sevenmsb
\scriptscriptfont\msbfam=\fivemsb
\def\Bbb#1{\fam\msbfam\relax#1}
\let\nd\noindent 

\def\natural{{\rm I\kern-.18em N}}

\def\mylesssim{\lower.8ex\hbox{$\ \mathop{\buildrel {\textstyle <}\over 
\sim}\nolimits\ $}}
\def\mygtrsim{\lower.8ex\hbox{$\ \mathop{\buildrel {\textstyle >}\over 
\sim}\nolimits\ $}}
\def\mysmalllesssim{\lower.8ex\hbox{$\mathop{\buildrel {\scriptstyle 
<}\over {\scriptstyle\sim}}\nolimits$}}
\def\mysmallgtrsim{\lower.8ex\hbox{$\mathop{\buildrel {\scriptstyle 
 >}\over {\scriptstyle\sim}}\nolimits$}}
\def\chix{{\raise.5ex\hbox{$\chi$}}}
\def\chixa{{\chix\lower.2em\hbox{$_A$}}}

\def\real{{\rm I\kern-.2em R}}
\def\integer{{\rm Z\kern-.32em Z}}
\def\complex{\kern.1em{\raise.47ex\hbox{
            $\scriptscriptstyle |$}}\kern-.40em{\rm C}}
\def\vs#1 {\vskip#1truein}
\def\hs#1 {\hskip#1truein}

\def\Month{\ifcase\number\month \relax\or January \or February \or
  March \or April \or May \or June \or July \or August \or September
  \or October \or November \or December \else \relax\fi }
\def\date{\Month \the\day, \the\year}

  \hsize=6truein        \hoffset=.25truein 
  \vsize=8.8truein      
  \pageno=1     \baselineskip=12pt
  \parskip=0 pt         \parindent=20pt
  \overfullrule=0pt     \lineskip=0pt   \lineskiplimit=0pt
  \hbadness=10000 \vbadness=10000 
     \pageno=0
     
     \footline{\ifnum\pageno=0\hss\else\hss\tenrm\folio\hss\fi}
     \hbox{}
     \vskip 1truein\centerline{{\bf Rigidity in Solids}}
     \vskip .2truein\centerline{by}
     \vskip .2truein
\centerline{{David Aristoff}
\ \ and\ \  {Charles Radin}
\footnote{*}{Research supported in part by NSF Grant DMS-0700120\hfil}}

\vskip .1truein
\centerline{ Mathematics Department, University of Texas, Austin, TX 78712} 
\vs.5 \centerline{{\bf Abstract}} 
\vs.2 \nd
We address the question of whether solids may be distinguished from
fluids by their response to shear stress.
\vs2
\vs1

     \vfill\eject

\nd {\bf I. Introduction}
\vs.1
Our focus is the theoretical modeling, within statistical mechanics,
of the solid/fluid phase transition of matter in thermal equilibrium,
for instance the 
\vs0 \nd ice/water transition, at high pressure and
temperature, and in particular the use of rigidity to distinguish the
phases.

There are no analytic proofs of a solid/fluid transition in any
statistical mechanics model which uses particles in space interacting
through simple short range forces (see however [1]), though there are
many simulations showing the transition, both Monte Carlo and molecular
dynamics. Since we concentrate on the transition at high pressure and
temperature, at which short range repulsion dominates the
interparticle interaction, the classical hard sphere model is
appropriate [2, page 84]; again there are no proofs of a phase transition in
this model, but there are many simulations [2,3].
Traditionally such a transition is ``understood'' theoretically through
an order parameter associated with some global (emergent) property of
the bulk material, in particular the molecular-level crystalline
symmetry which can be detected experimentally in X-ray scattering
[4]. This paper follows an alternative proposal of Anderson [5],
namely the use of rigidity, the response to stress, to distinguish the
phases, for instance in a hard sphere model.

Stress, both pressure and shear, will be understood here as an
external influence (force per unit area) on the boundary of a finite
sample of the material, with pressure acting on the volume and shear on the
shape; its extension to a uniform stress field inside the material is
an important issue to be addressed.
Pressure is commonly modeled in statistical mechanics
as a parameter in a pressure ensemble which controls the deformations of
volume. A similar approach can be followed for shear stress, using a set of
parameters controlling fluctuations of the shape of the container,
though this is much less common; see however [6,7].

One might in principle be able to model the distinction between ice and water through
statistical mechanics estimates of compressibility ($-[\partial
V/\partial p]/V$), in
which $V$ is volume and $p$ is pressure. Unfortunately the
compressibility of ice and water are both high and the difference is
relatively small, a common circumstance for a solid/fluid
transition. However the difference in the corresponding elastic shear 
constants is, experimentally, much greater, since elastic shear constants are
negligible for (isotropic) fluids. This suggests an advantage in using
shear instead of pressure to distinguish a solid from its melt.

But, as emphasized for instance in the recent paper of
Sausset et al [8], for a material in equilibrium any linear response 
to macroscopic shear must be transient in
time, making it harder to model an elastic shear constant within
equilibrium statistical mechanics. Indeed, there are proofs that
in equilibrium statistical mechanics the free energy is independent of
the shape of the material [9,10], which suggests that shear stress be properly
considered as taking a material out of equilibrium. 
In [8] solids are treated as highly viscous fluids and solids
are distinguished from fluids by a dynamical feature, the divergence of the viscosity as
the shear stress rate vanishes. Since we are using an equilibrium
model we focus instead on a spatial issue, namely
the question of whether the
response to shear is localized at a small part of the material or is (uniformly)
distributed throughout the material. We concentrate on the
response to shear stress in the limit of zero shear,
computed before any bulk limit. (At zero shear stress the shape
of the material is unconstrained; imagine a triaxial shear cell, with
negligible friction and in zero gravity.) Mathematically, we are
taking advantage of the fact that the limit of vanishing shear need
not commute with the infinite volume limit. 

Our approach is based on the following idea. A
configuration of hard spheres at high pressure must be approximately
crystalline, with most particles trapped in a cage of neighbors. A
macroscopic change in container shape can be accommodated by
adjustments only
near the boundary, without affecting the structure in the bulk
interior; see Fig.\ 1. On the contrary a very small change of shape
cannot be accommodated in this way and might result 
in a small deformation of internal structure throughout the
configuration rather than just near the boundary; see Fig.\ 2. 

\epsfig .8\hsize; 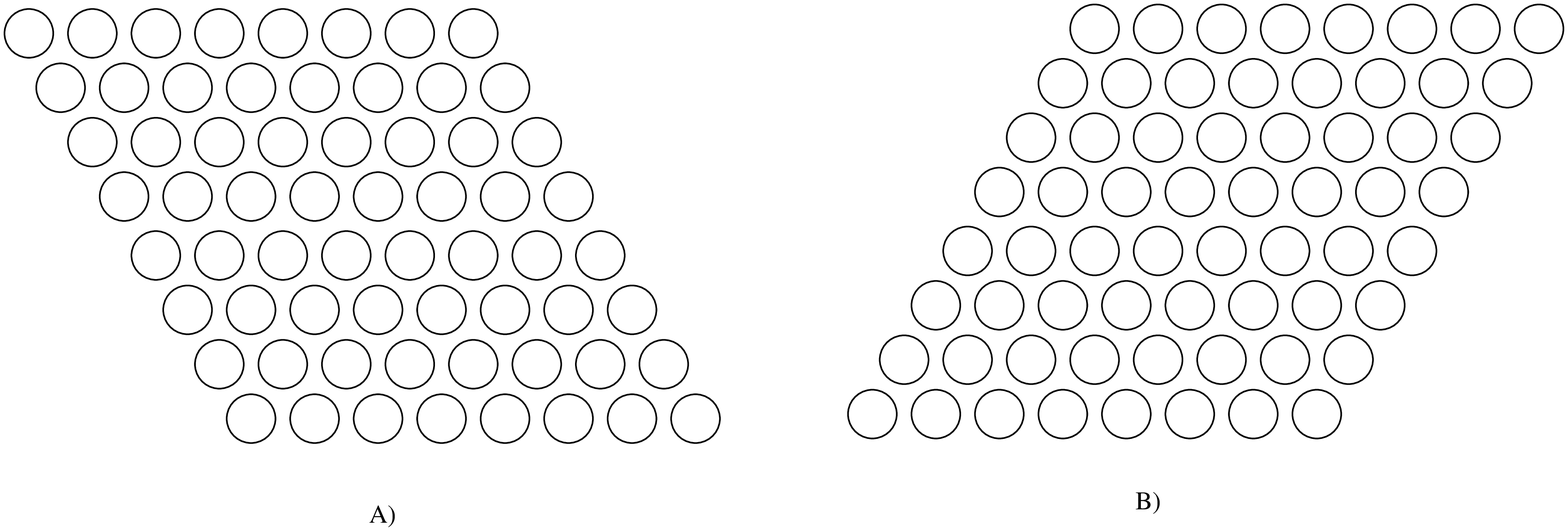;
\vs-.2 \nd
{\bf Figure 1.} The result of a macroscopic change of strain: the interiors of A) and
B) are the same.
\vs0
\epsfig .8\hsize; 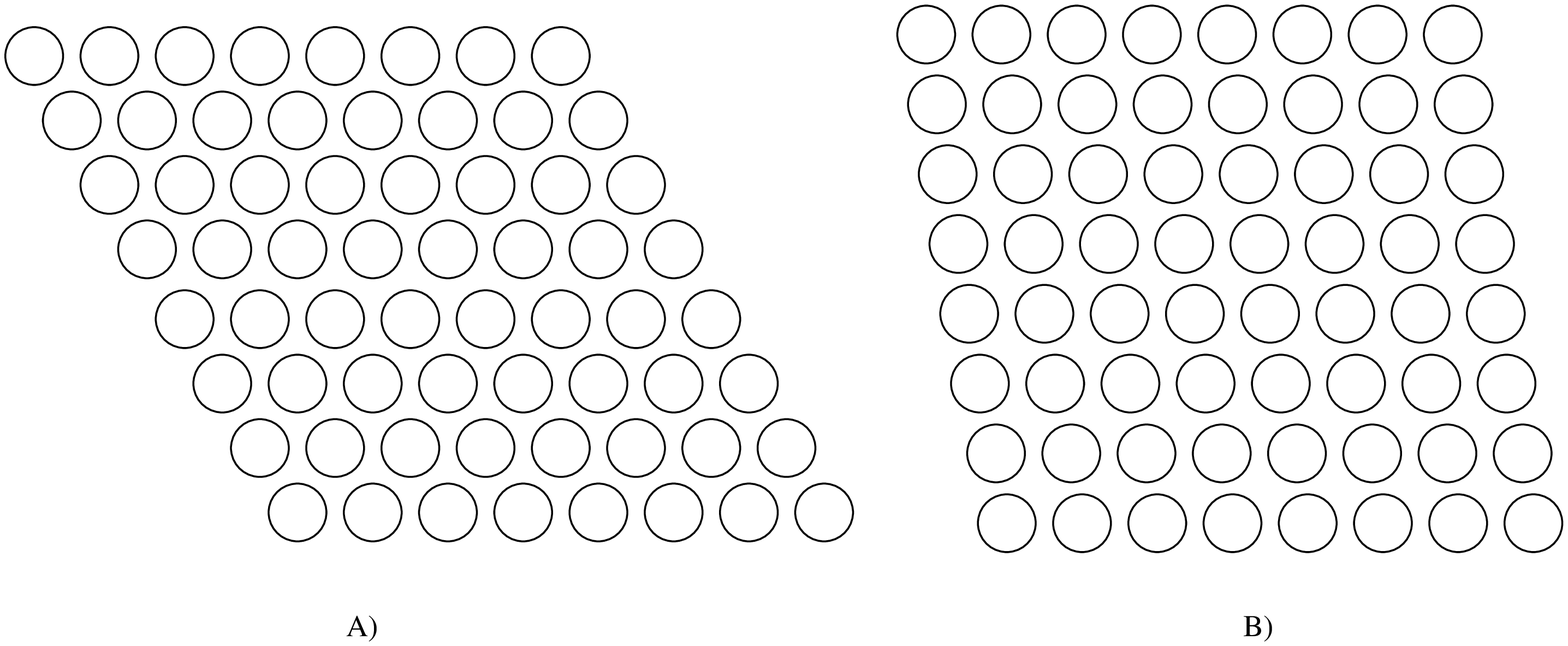;
\vs-.2 \nd
{\bf Figure 2.} The result of a small change of strain: the interiors of A) and
B) are different, the underlying lattice becoming distorted.
\vs.1
In
other words, there may be a regime in which the response to
shear is linear and extends throughout the material, but such a 
regime, measured by the angle of
deformation, would have to vanish with the size of the system,
constituting an equilibrium alternative to the dynamical effect discussed
by Sausset et al.
If indeed the response extends throughout the material, which is by no
means evident,
it would be appropriate to measure it, in a finite system at constant high
pressure, by the rate of change in density $\phi$ with respect to shear stress $f$,
computed at $f=0$. We expect this to be large in magnitude at high pressure,
in the solid phase. At low pressure the model should represent a fluid
with negligible response, and we propose this difference as a means of
detecting the solid/fluid phase transition. 

In the remainder of this paper we give indirect support to the proposition that
for infinitesimal shear the material responds linearly throughout the finite sample.
We do this by simulation of the two dimensional hard disk model, in a
stress (pressure and shear) ensemble. Our measurements in this model 
show an emerging resistance to shear at volume fraction about $0.7$,
very close to the known transition(s) for hard disks. Our
simulation does not show the details of the transition, which are well
displayed in the recent tour-de-force by Bernard and Krauth
[11]. Instead, the point of this work is merely to illustrate the
feasibility of using density response to infinitesimal shear to probe
a solid/fluid phase transition, in the tradition of Anderson [5].
\vs.2
\nd {\bf II. The Model and Simulations}
\vs.1
We consider arrangements of a fixed number $N$ of hard disks of fixed radius $\sigma$ inside various parallelograms,
with the volume and shape of the parallelograms allowed to vary. More 
precisely, we consider arrangements of such disks inside boundaries 
formed by placing disks along the edges of parallelograms at regular 
intervals as in Fig.\ 3. These boundary disks lie on a regular 
triangular lattice when the underlying parallelogram is rectangular, and all 
the parallelograms are related to each other via maps (on the 
boundary disk centers) of the form 
$(x,y) \to (\lambda x + \nu \lambda y, \lambda y)$ for real $\lambda, \nu$, with $\lambda > 0$.

We employ  
a ``stress ensemble'' which uses parameters $p$ and $f$ 
to control the volume $V$ and deformations of shape $\alpha$, respectively, of the parallelograms. 
(We want to use the Lagrange multipliers $p$ and $f$ to model pressure
and shear stress.) More precisely, we 
consider probability measures (states) which minimize the free energy 
$$F(p,f) = S-\beta pV+\beta f\alpha V.\eqno(1)$$
Here $\beta$ is the inverse temperature and $\alpha$ is the angle of inclination of a parallelogram, 
with $\alpha = 0$ representing a rectangle (see Fig.\ 3). Such an ensemble 
has partition function 
$$Z_{p,f} = \int_0^\infty \int_0^\infty \left(\int_{V,\alpha} dC\right) \exp(-\beta pV+\beta f\alpha V) dV d\alpha,\eqno(2)$$
where $\int_{V,\alpha} \ dC$ represents integration over all arrangements of hard disks in a parallelogram of 
volume $V$ and shape $\alpha$. (Temperature plays a simple role in
hard core models such as this, so we will assume that numerically
$\beta =1$ where convenient. Also we are using the usual ``reduced''
ensemble in which the velocity variables have been integrated out.)
By the change of coordinates 
$$\psi_{V,\alpha}: (x,y) \to (V^{-1/2}(x-\tan(\alpha)y),V^{-1/2}y),\eqno(3)$$
considered as a function on the disk centers, 
the partition function may be rewritten as 
$$Z_{p,f} = \int_0^\infty \int_0^\infty \int_{\Omega} \Phi(\psi_{V,\alpha}^{-1}(Q)) V^N \exp(-\beta pV+\beta f\alpha V) dQ dV d\alpha,\eqno(4)$$
where $\Phi(C) = 1$ if no two disks of radius $\sigma$ with centers from $C$ overlap, $\Phi(C) = 0$ otherwise, and 
$\Omega \subset {\Bbb R}^2$ is a fixed rectangular area. 
The probability density of an arrangement of (nonoverlapping) hard disks in a parallelogram of volume $V$ and shape $\alpha$ is then 
proportional to 
$$V^N \exp(-\beta pV+\beta f\alpha V).\eqno(5)$$
Because we are interested only in infinitesimal shear, 
we impose the restriction $0 \le \alpha \le 0.01$, with $\alpha$ measured in radians. 
(For the densities and $\alpha$ we consider, the boundary disks do not come close to 
overlapping.)

\vs.1 \nd
\epsfig .7\hsize;  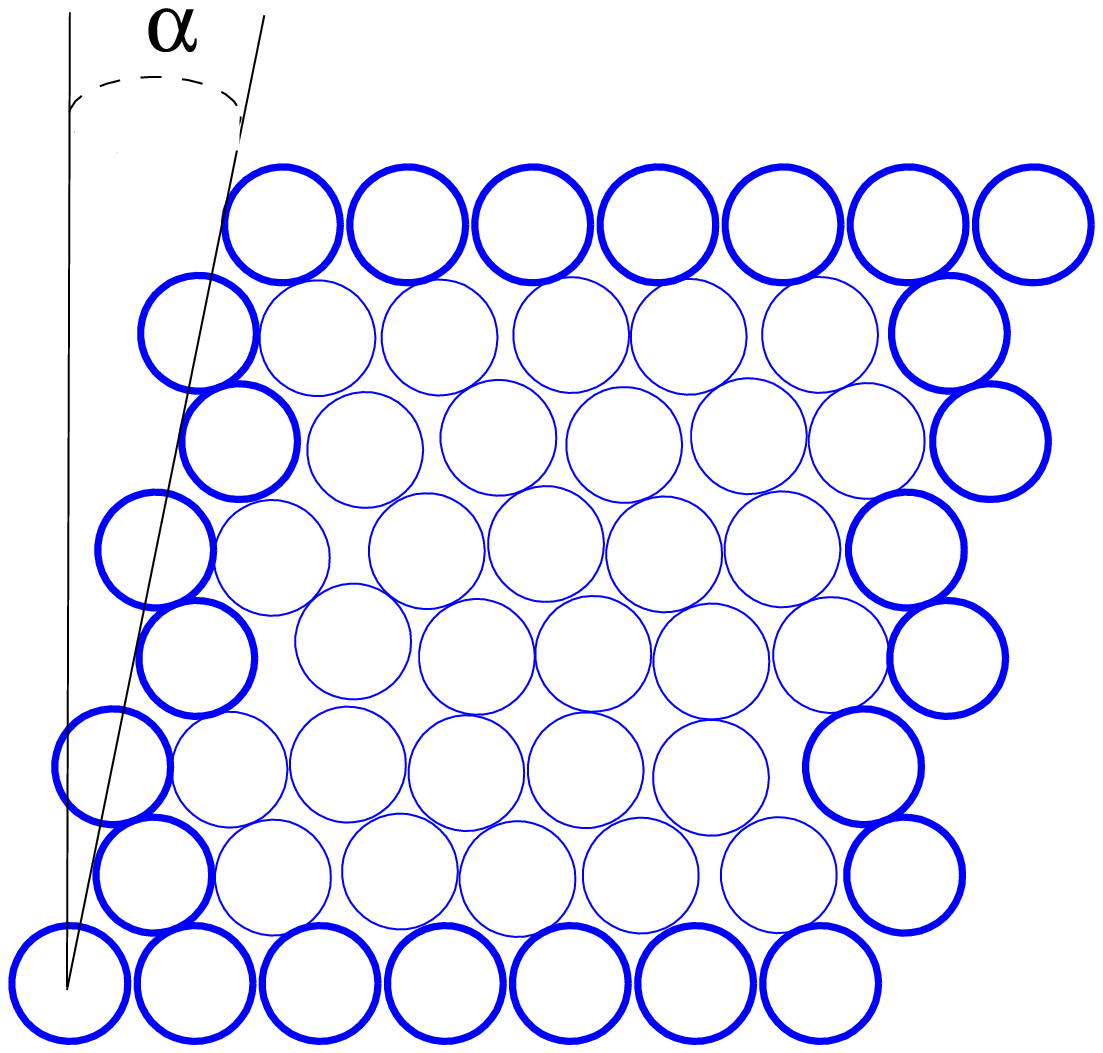;
\vs-.2 \nd
{{\bf Figure 3.} An arrangement of disks in a parallelogram. 
Boundary disks are in bold;}
{$\alpha$ is the angle formed between the boundary disks and a vertical line.}
\vs.2 \nd

Let $\phi_{p,f}$ be the average volume fraction of arrangements at fixed $p$ and $f$. 
To measure the volume response of arrangements of disks to an infinitesimal change in shape, 
we estimate the derivative 
$$\Gamma(p) := {{\partial \phi_{p,f}}\over{\partial f}}\bigg|_{f=0}.\eqno(6)$$ 
By definition the average volume fraction $\phi_{p,f}$ is given by 
$$\phi_{p,f} = \int_0^\infty \int_0^\infty \int_{\Omega} \Phi(\psi_{V,\alpha}^{-1}(Q)) (N\pi\sigma^2/V) V^N \exp(-\beta pV+\beta f\alpha V) dQ dV d\alpha.\eqno(7)$$
Differentiating with respect to $f$, one obtains 
$$\Gamma(p) = \beta N\pi\sigma^2 [\langle \alpha \rangle_{p,0} - \Big\langle {1\over V}\Big\rangle_{p,0}\langle V\alpha \rangle_{p,0}]\eqno(8)$$
with $\langle \cdot \rangle_{p,f}$ representing an average value at fixed $p$ and $f$. 
Applying equation (8), we obtain $\Gamma(p)$ from the average values
of $\alpha$, $1/V$ and $V\alpha$, in our simulations at pressure $p$ with $f = 0$. 

\vs.1 \nd
\epsfig 1\hsize;  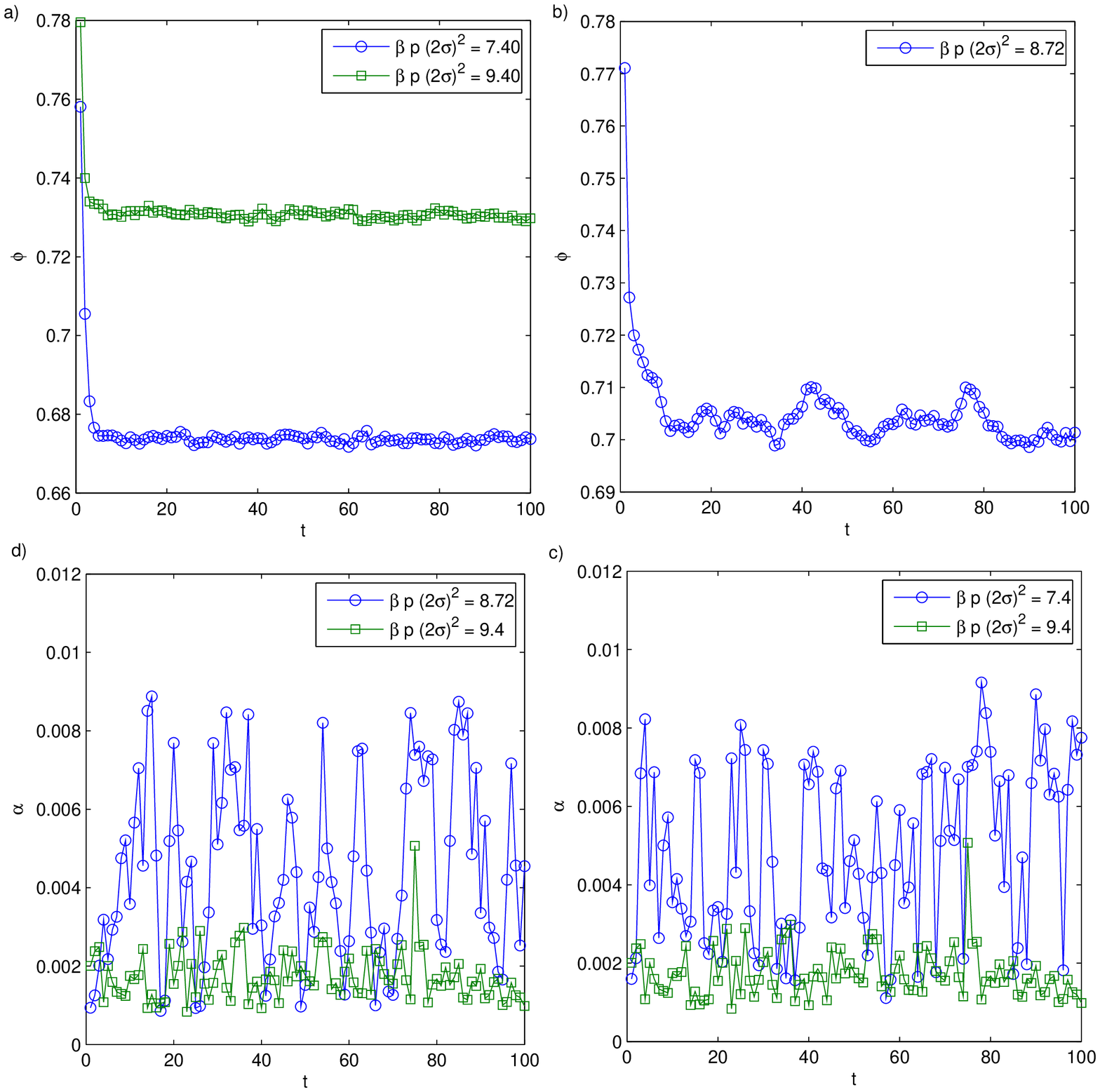;
\vs-.2 \nd
{{\bf Figure 4.} Time series for systems of $N=10656$ disks,  
with $t$ in units of $3.5\times 10^8$ Monte Carlo steps, and 
at $f = 0$, clockwise from 
top left: a,b) Volume fraction $\phi$ 
vs. $t$; c,d) shape $\alpha$ vs. $t$.}
\vs.1 \nd

The change from $\Gamma(p) \equiv 0$ to $\Gamma(p) << 0$ represents 
the development of rigidity in the system, in the sense that the 
system has to overcome pressure $p$ to expand in response to infinitesimal shear.


Our simulations use Monte Carlo steps consisting of moves which 
change the size and shape of the arrangements of disks, as well as 
so-called ``event-chain'' movements of multiple disks [11]. In the former types 
of moves, the coordinates of disk centers $(x,y)$ are mapped to 
$(\lambda x, \lambda y)$ or $(x + \rho y, y)$, respectively. Here 
$\lambda = (V+ \eta \epsilon)^{1/2}/V^{1/2}$ and $\rho = \nu \delta$, 
where $\eta$, $\nu$ are (independent) random variables distributed uniformly 
in $(-1,1)$, and $\epsilon$ and $\delta$ are positive real parameters. 

\vs.1 \nd
\epsfig 1\hsize;  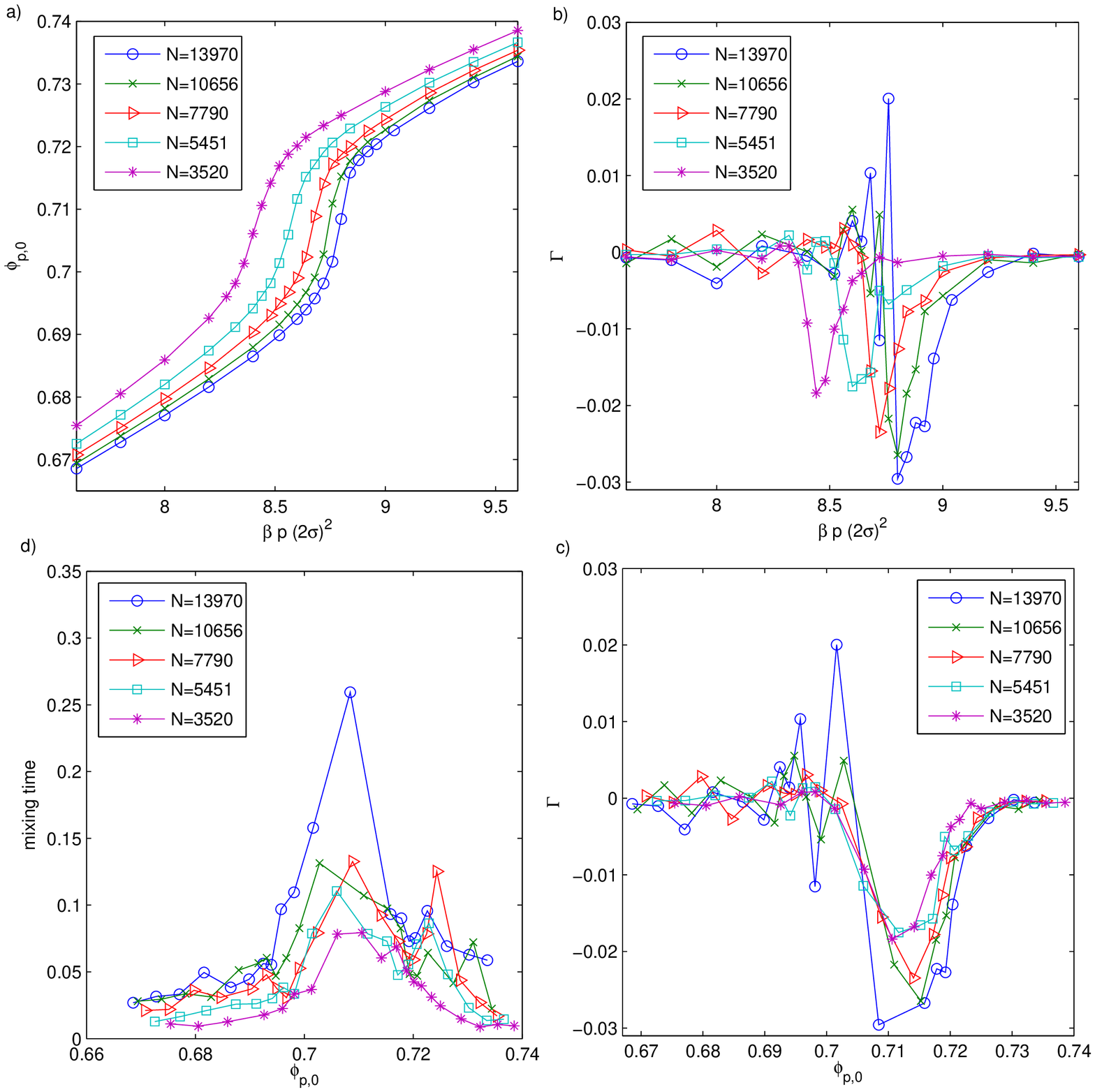;
\vs-.2 \nd
{{\bf Figure 5.} Volume fraction data, clockwise from top left: a) Average volume fraction $\phi_{p,0}$ vs. pressure 
$\beta p (2\sigma)^2$, with confidence intervals smaller than the data markers; 
b) Differential volume response $\Gamma$ vs. pressure $\beta p (2\sigma)^2$; 
c) Differential volume response $\Gamma$ vs. average volume fraction $\phi_{p,0}$; d) Mixing time, 
as a fraction of total number of Monte Carlo steps, vs. $\phi_{p,0}.$}
\vs.2 \nd
If such a move results in overlap of disks, it is rejected. These 
types of moves are (each) attempted with frequency $N^{-1/2}/4$.
In the latter type of move, employed recently in [11], a non-boundary 
disk and a random direction are selected, 
with the direction being up, down, left or right (that is, 
parallel to one of the coordinate axes). 
Additionally a displacement $L$ is selected uniformly at random from 
the interval $(0,\sqrt{V-N\pi\sigma^2}/2)$. The particle is then 
moved along the chosen direction 
\vs.1 \nd
\epsfig 1\hsize;  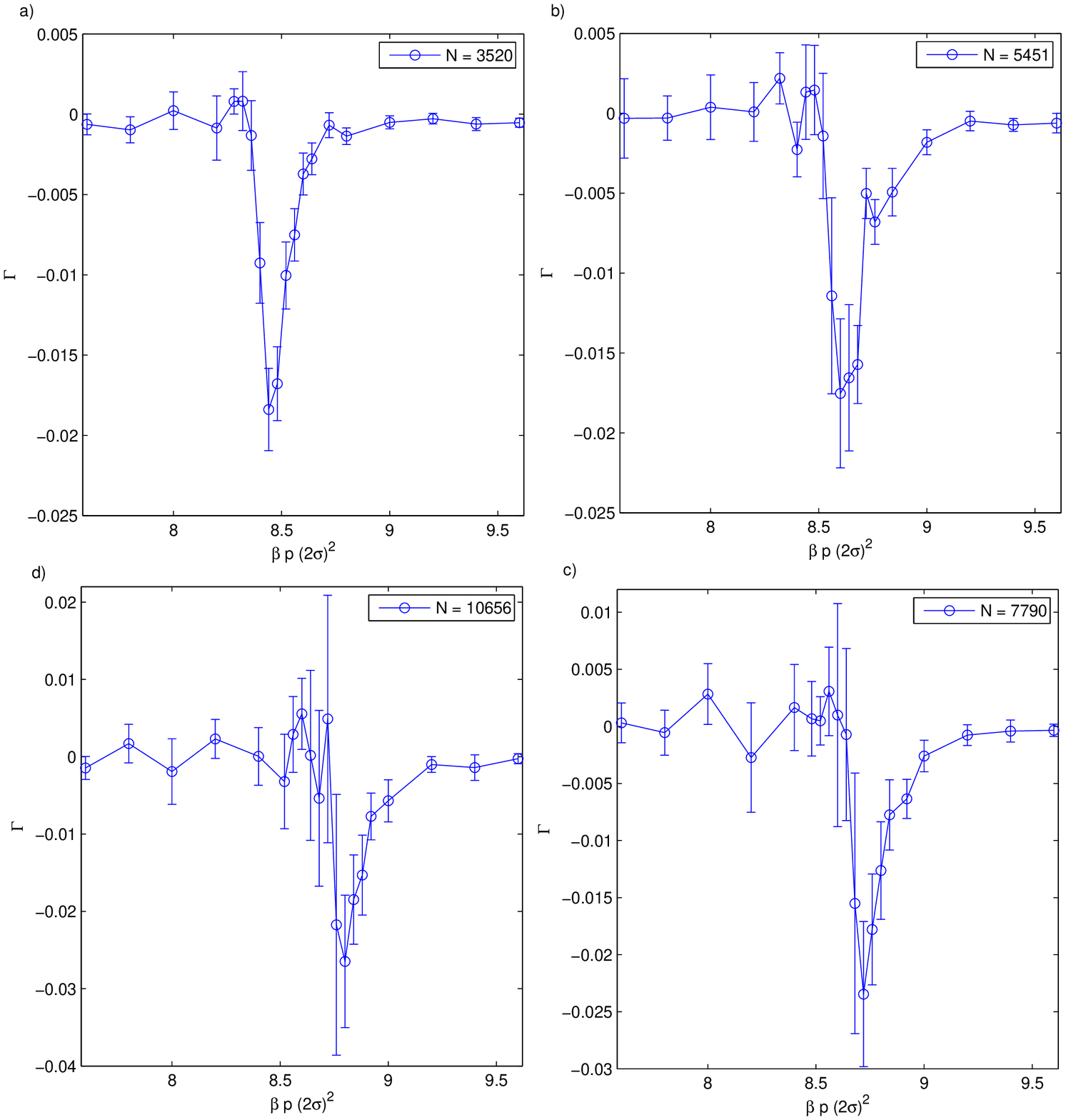;
\vs-.2 \nd
{{\bf Figure 6.} Confidence intervals for differential volume response $\Gamma$  vs. pressure 
{$\beta p (2\sigma)^2$, for systems with (clockwise from top left) $N=3520$, $N=5451$, 
$N=7790$, and $N=10656$.}
\vs.2 \nd
until it strikes another particle, at 
which point that particle moves in the same direction until it strikes 
another particle, and so on. The process continues until a total displacement of $L$ 
is obtained, the total displacement being the sum of the displacements of all the 
particles. 
If the process results in the displacement of a boundary particle, 
then the move is rejected. (It is in principle possible that such moves result in disks 
moving outside the boundary, but this does not occur for the pressures we 
simulate.) Such moves are attempted with frequency $1-N^{-1/2}/2$.

We investigate systems with $N=3520$, $N=5451$, $N=7790$, $N=10656$, and $N=13970$ disks, 
beginning with perfectly crystalline arrangements of the disks. 
At each pressure $p$ we run these systems to $2\times 10^{10}$, $2.5\times 10^{10}$, 
$3\times 10^{10}$, $3.5\times 10^{10}$, and $4\times 10^{10}$ Monte Carlo steps, respectively. 
This results in about $2\times 10^7$ displacements per particle, and about $10^7$ fluctuations 
in volume and shape, for each $p$ and 
system size $N$. We checked that our runs were long enough for volume fraction $\phi$ 
and shape $\alpha$ to equilibrate, after a burn-in time of at most about $10\%$ of 
the run (with the exception of the shape $\alpha$ at large $p$, as discussed below); 
see Fig.\ 4.  Therefore in our main data, shown in Fig.\ 5 (with confidence intervals in Fig.\ 6), 
we have thrown away the first $10\%$ of each Monte Carlo run. Along with our main data we 
also measured mixing times, defined as the number of Monte Carlo steps before the 
standard (unbiased) autocorrelation of the time series for $\phi_{p,0}$ first crosses zero; see Fig.\ 5. 
Excluding the largest system ($N = 13970$), mixing times were no more than $15\%$ of our 
Monte Carlo runs. For $90\%$ confidence intervals, we run $10$ independent copies of every simulation 
and use Student's $t$-distribution with $9$ degrees of freedom on the average values obtained from each of the $10$ copies; 
see Fig.\ 6. We do not compute confidence intervals for the largest system, $N = 13970$.

We find the volume response parameter $\Gamma(p)$ defined in (8)
exhibits the following behavior (see Fig.\ 5). 
At low pressure $p$ or volume fraction $\phi = \phi_{p,0}$, $\Gamma(p)$ is approximately zero, indicating 
there is no volume response to an infinitesimal shear. We interpret this as meaning 
the hard disk fluid does not resist a small 
shear stress. As $\phi$ rises above $0.70$, the volume response $\Gamma$ begins to 
deviate from zero. Our data is not fine enough to distinguish the details
of the transitions shown in [11], and in particular does not justify
estimating specific transition values for $p$.



We note that the volume response $\Gamma$ measured in our
simulations tapers off to zero at large $p$ (or high density).
We do not expect this tapering to be accurate; instead we 
interpret this as a sign that the simulations begin to get ``stuck'' as densities increase. 
This is confirmed by checking that 
the simulations no longer explore the full range of shapes $\alpha$; see Fig.\ 4c) -d). 
We expect the true behavior of $\Gamma$, as a function of $\phi$, to be non-increasing, 
indicating a volume response into the nonfluid phases. 
We conclude, then, that the hard disk solid resists a 
small positive shear stress, while the fluid does not.
\vs.2
\nd {\bf III. Summary}
\vs.1
The rigidity of solids can be modeled in various ways. We have chosen
to use equilibrium statistical mechanics, in large extent because it
is the most convincing formalism in which to distinguish solids from
fluids, which is our motivation for studying rigidity, following Anderson
[5]. Even within equilibrium statistical mechanics one could 
model response to shear more
simply with a harmonic crystal model [12: Chapter 22], with long range quadratic forces
between particles assigned neighboring equilibrium sites. In fact this
is commonly used to
model sound (pressure) propagation, but does not exhibit
a fluid phase and therefore does not allow comparison between solid
and fluid phases, which is the purpose of our work.

The most awkward consequence of using equilibrium statistical
mechanics is that to obtain the sharp solid/fluid distinction one must
take the infinite volume limit 
while for infinite systems one cannot model shear stress, as noted
in the introduction. Our solution to this was to model
the shear in finite volume -- where there are no well defined solid or
fluid phases but the system can model shear stress -- 
and measure the volume response in
the limit of vanishing shear, before taking the infinite volume
limit. 

The other weakness of our approach is technical: in order to measure
the volume response to stress we employed variation in
both strain and volume, which is costly in simulation time compared to
the usual Monte Carlo technique for the hard disk system, which 
uses fixed volume, strain and particle number. We favor this ensemble 
for its theoretical advantages: calculating 
response to stress in our ensemble is far simpler than in an 
ensemble which fixes, say, density and strain -- in our 
ensemble we can calculate the density response directly from fluctuations, whereas 
an equivalent analysis in the latter model would require taking a 
numerical derivative of average pressure as shear strain vanishes, with pressure 
computed approximately (perhaps from a virial expansion; see [11]). 
Our data, due to the large computation time associated with fluctuations in volume and strain, 
 is not sufficient to demonstrate the details of the transitions, 
as is done in [11]. We feel this is acceptable in exchange for demonstrating 
the feasibility of shear response as a theoretical tool to analyze the solid/fluid transition. 

In conclusion we note that our approach is similar to the analysis of 
the dilatancy transition
recently found experimentally [13] in (nonequilibrium) static,
granular matter, and its modeling [14] with a stress ensemble. In
effect we are proposing to model the solid/fluid transition of
equilibrium matter as a dilatancy transition, a sharp change between the
solid and fluid equilibrium phases in their volume response to (infinitesimal) shear, instead of by a
change in symmetry of the molecular pattern, as is the common
practice.
\vs.2 \nd
{\bf Acknowledgements.} It is a pleasure to acknowledge useful
discussions with Giulio Biroli, Daan Frenkel and Dan Stein.

\vfill \eject
\centerline{Bibliography}
\vs.1 \nd

\item{[1]} L. Bowen et al, Fluid-solid transition in a hard-core
system, Phys. Rev. Lett. 96(2006), 025701.

\item{[2]} B. J. Alder, W. G. Hoover, Numerical statistical mechanics,
    in Physics of Simple Liquids, edited by H. N. V. Temperley,
    J. S. Rowlinson and G. S. Rushbrooke, John Wiley, New York, 1968,
    79-113.

\item{[3]} H. L\"owen,
{ Statistical Physics and Spatial Statistics: The art of analyzing 
spatial structures and pattern formation}, ed. K. Mecke and D. Stoyan,
Springer, 2000.

\item{[4]} J.M. Yeomans, Statistical Mechanics of Phase Transitions,
Clarendon Press, Oxford, 1992.

\item{[5]} P. W. Anderson, Basic Notions of Condensed Matter Physics,
    Benjamin/Cummings, Menlo Park, 1984, Chapter 2.

\item{[6]} M. Parrinello and A. Rahman, 
Strain fluctuations and elastic constants,
J. Chem. Phys. {76}(1982), 2662.

\item{[7]} I.R. MacDonald,
NpT-ensemble Monte Carlo calculations for binary liquid mixtures, 
Molecular Physics Vol {100}(2002), 95.

\item{[8]} F. Sausset, G. Biroli and J. Kurchan,
Do solids flow?,
J. Stat. Phys. 140(2010), 718-727.

\item{[9]} J. Lebowitz,
Statistical mechanics -- a review of selected rigorous results, 
Ann. Rev. Phys. Chem. 19(1968) 389-418.

\item{[10]} D. Ruelle, { Statistical Mechanics; Rigorous Results}, Benjamin, New
York, 1969.

\item{[11]} E.P. Bernard and W. Krauth, First-order liquid-hexatic phase
transition in hard disks, arXiv:1102.4094.

\item{[12]} N.W. Ashcroft and N.D. Mermin, Solid State Physics,
Saunders College, Philadelphia, 1976.

\item{[13]} J.-F. M\'etayer et al, Shearing of frictional sphere
packings, EPL 93(2011), 64003.

\item{[14]} D. Aristoff and C. Radin, Dilatancy transition in a granular model,
J. Stat. Phys. 143(2011), 215-225.

\end